\documentclass[twocolumn,amsmath,amssymb]{snp}
\usepackage{graphicx}
\usepackage{dcolumn}
\usepackage{bm}
\begin{document}

\title{{\Large Quarkonia production at forward rapidities with the ALICE Experiment}}
\author{\large Debasish Das$^{1*}$}
\affiliation{$^1$Saha Institute of Nuclear Physics, 1/AF Bidhannagar, Kolkata 700064, INDIA.\\
*email: debasish.das@saha.ac.in, debasish.das@cern.ch\\
(For the ALICE Collaboration)}

\begin{abstract}
\leftskip1.0cm
\rightskip1.0cm

The suppression pattern of quarkonia in
AA(nucleus-nucleus) collisions (where the plasma of 
deconfined quarks and gluons, the Quark-Gluon Plasma (QGP) formation is expected), 
along with the comparative quarkonia production in pp collisions provides important
understanding into the properties of the produced medium. 
Experimental results of Relativistic Heavy Ion Collider(RHIC) indicate 
a suppression on charmonium production in AA collisions. Muons from the decay of charmonium
resonances are detected in the ALICE Experiment at 
the Large Hadron Collider(LHC) for pp and Pb-Pb collisions with
a muon spectrometer, covering the forward rapidity region
2.5$<$ $y$ $<$4.0. Analysis of the nuclear modification factor ($R_{\rm AA}$) at forward rapidity 
are presented and compared with mid-rapidity results from 
electrons in the central barrel covering $|y|<$0.9. 
The roles of suppression and regeneration mechanisms 
are discussed, as well as the importance of the results of the 
forthcoming p-Pb data taking for the estimate of cold nuclear matter 
effects on quarkonia. Perspectives for the bottomonia measurements are also given.
Quarkonia results via muon channel from CMS experiment at LHC are
compared with ALICE quarkonia measurements.

\end{abstract}\maketitle

\section{Introduction}

Quantum Chromo-Dynamics(QCD) is the well-established theory of strong interactions.
Quarks are the basic constituents of QCD, which interact through the exchange of gluons. 
A thermalized system where the properties of the system are governed by the quarks and gluons
degrees of freedom is called the ${\it Quark~-Gluon~Plasma}(QGP)$~\cite{Shuryak:1980tp}. 
Lattice QCD predicts this new phase, QGP, 
at high temperatures($\sim$155-160 MeV)~\cite{Bazavov:2011nk,Aoki:2009sc}.
The motivation of the relativistic heavy ion physics is the experimental study of the 
hadronic matter under extreme conditions of temperature~\cite{Harris:1996zx}.
This is the main reason for the present heavy ion program ongoing 
at the Relativistic Heavy-Ion Collider (RHIC, at Brookhaven National Laboratory in New 
York) and at the Large Hadron Collider (LHC, at the European Organization 
for Nuclear Research in Geneva).

A wealth of ideas have been proposed in the past few decades on the 
experimental and theoretical understanding 
of QGP properties~\cite{Muller:2006ee,Muller:2012zq,Blaizot:2007sw,Bass:1998vz}. 
The plasma is short lived and all signals from QGP have background 
from the hot hadronic phase that follows the QCD phase transition. 
Therefore, it is important to understand the physics in hot hadronic phase with different 
experimental probes~\cite{Harris:1996zx}.

\section{Quarkonia in heavy-ion collisions}

Among the possible probes of the QGP, the interest in heavy quarks
is motivated by their unique role in the diagnostics of the highly 
excited medium created in relativistic heavy-ion collisions. 
Both experimentally and theoretically for over two decades, 
the properties of heavy quarkonium states(which are bound states of heavy quark-antiquark pairs, 
charmonium and bottomonium) in a hot and dense QCD medium have 
been intensely studied~\cite{Kluberg:2005yh,Brambilla:2004wf}. 
Heavy-quark mass is large and so heavy-quark production is 
believed to be occur largely within the earliest phase of the collision. Therefore, the measurement of 
quarkonia is expected to provide essential information on the properties of the strongly-interacting 
system formed in the early stages of heavy-ion collisions~\cite{Eichten:1979ms}.

If a $J/$$\psi$ particle(a $c\overline{c}$ bound state) is placed in QGP, the color
charge of the charm quark $c$ will get screened  by the quarks, anti-quarks and the gluons of the plasma.
The charmonium production was discussed as an important probe of plasma~\cite{Shuryak:1978ij,Cleymans:1984zs}, along with 
the suggestion by Matsui and Satz~\cite{Matsui:1986dk} which affirmed that with increasing 
centrality in heavy-ion collisions a suppression of the $J/$$\psi$ is expected. 
The dissociation of the charmonium 
bound-state in a dense medium is connected with the Debye screening 
of the binding potential. This is influenced by a deconfinement of color 
charges and thus intimately connected to the formation of a 
QGP~\cite{Blaschke:2011zz,Gerschel:1998zi,Vogt:1999cu,Satz:2005hx,Kharzeev:2007ej,Bali:2000gf,Kaczmarek:2005ui}.

Comprehensive experimental results at SPS~\cite{Kluberg:2005yh,Abreu:2000ni} (including feeddown from 
other less bound resonances like $\psi(2S)$  and $\chi_{c}$) and RHIC of  
$J/$$\psi$ production in AA collisions clearly indicate that even the 
strongly bound $J/$$\psi$ ground state is suppressed~\cite{physreptvogt1,Adler:2004ta,Abelev:2009qaa,Adare:2006ns}. 
However at LHC energies the $J/$$\psi$ 
production could be even enhanced due to the coalescence 
of uncorrelated $c\bar{c}$ pairs in the medium~\cite{BraunMunzinger:2000px}.
Initial state effects like  modifications of the parton distribution functions in 
the nucleus relative to the nucleon shadowing need to be taken into account. 
The final state effects like the nuclear absorption are expected 
to be practically insignificant at the LHC energies. 
Various statistical and 
transport models~\cite{BraunMunzinger:2000px,Andronic:2011yq,Zhao:2011cv,Liu:2009nb} are proposed for LHC energies.
Studying the pA collisions at the LHC energies is crucial to quantify 
the role of initial shadowing effects.

\subsection{$J/$$\psi$ $R_{\rm AA}$}

A Large Ion Collider Experiment (ALICE)~\cite{jinst} is a general-purpose heavy-ion experiment 
to study the nature of the quark matter under extreme temperature ($\geq$ 0.3 GeV)~\cite{physperf1,physperf2} 
created in nucleus-nucleus collisions at the LHC. In the framework of the ALICE physics program, 
the goal of the Muon spectrometer is the study of quarkonia 
production~\cite{Aamodt:2011gj,Abelev:2012kr,alicePbPb2010}, along with 
open heavy flavor production and low mass vector meson properties~\cite{ALICE:2011ad} 
via the di-muon decay channel in pp, pA and AA collisions. Using the particle identification 
potential of the central barrel detectors($|\eta|<$0.9) $J/$$\psi$ have been also 
measured in di-electron channel at mid-rapidity~\cite{Aamodt:2011gj,Abelev:2012kr}.

\begin{figure}
\includegraphics[scale=0.39]{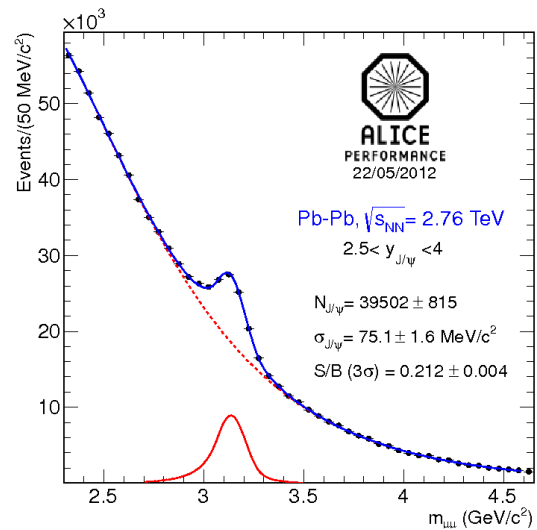}
\caption{\label{fig1}(Color Online) Opposite sign di-muon invariant mass distribution for the 0$\%$-90$\%$ 
most central collisions for 2.5$<y<$4, integrated over $p_{T}$.}
\end{figure}

\begin{figure}
\includegraphics[scale=0.365]{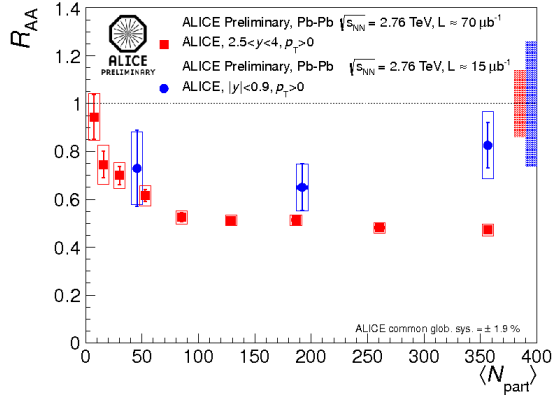}
\caption{\label{fig2}(Color Online) Inclusive $J/$$\psi$  $R_{\rm AA}$ for mid-(blue) and forward(red) rapidity as a 
function of the number of participating nucleons measured in 
Pb-Pb collisions at $\sqrt{s_{\rm NN}}$ = 2.76 TeV.}
\end{figure}

\begin{figure}
\includegraphics[scale=0.365]{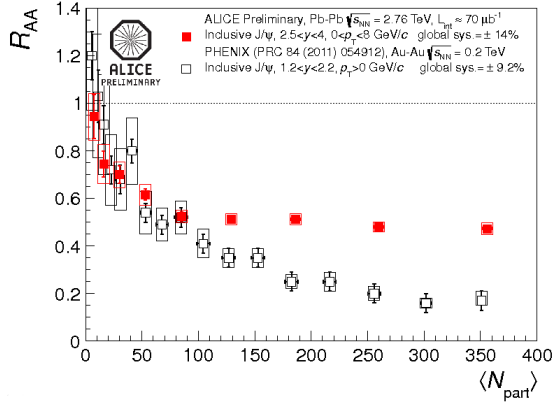}
\caption{\label{fig3}(Color Online) $J/$$\psi$  $R_{\rm AA}$ in Pb-Pb collisions at $\sqrt{s_{\rm NN}}$ = 2.76 TeV recorded in 2011 as a function of the number of participating nucleons and compared with PHENIX results~\cite{Adare:2011yf} in Au+Au collisions at $\sqrt{s_{\rm NN}}$ = 200 GeV.}
\end{figure}

\begin{figure}
\includegraphics[scale=0.365]{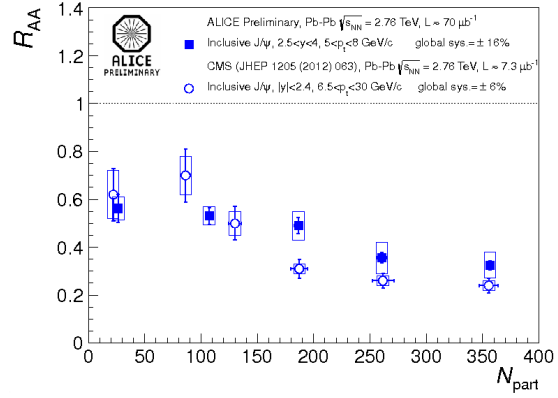}
\caption{\label{fig4}(Color online) $J/$$\psi$  $R_{\rm AA}$ for high-$p_{T}$(5$<$$p_{T}$$<$8 GeV/$c$) 
in Pb-Pb collisions at $\sqrt{s_{\rm NN}}$ = 2.76 TeV as a function of the number of 
participating nucleons and compared with CMS results~\cite{Chatrchyan:2012np}.}
\end{figure}

\begin{figure}
\includegraphics[scale=0.365]{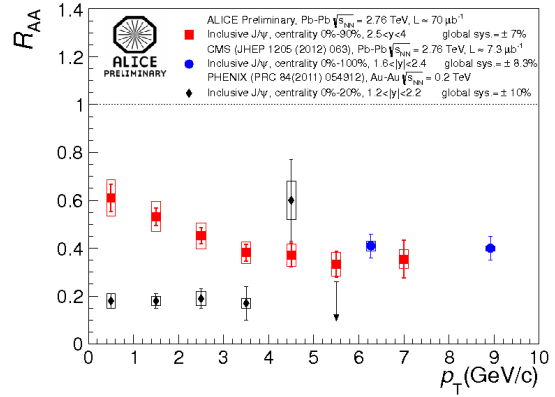}
\caption{\label{fig5}(Color online) Inclusive $J/$$\psi$  $R_{\rm AA}$ as a function of the $J/$$\psi$ $p_{T}$  for 2.5$<$$y$$<$4.0 and centrality 0$\%$-90$\%$ compared with CMS~\cite{Chatrchyan:2012np} and PHENIX~\cite{Adare:2011yf} data.}
\end{figure}

\begin{figure}
\includegraphics[scale=0.368]{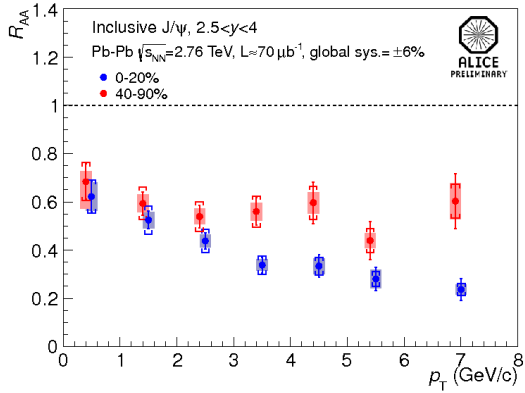}
\caption{\label{fig6}(Color online) Comparison of $R_{\rm AA}$ as a function of the $J/$$\psi$ $p_{T}$ 
for 2.5$<$$y$$<$4.0 in two centrality classes of 0$\%$-20$\%$ and 40$\%$-90$\%$.}
\end{figure}

\begin{figure}
\includegraphics[scale=0.365]{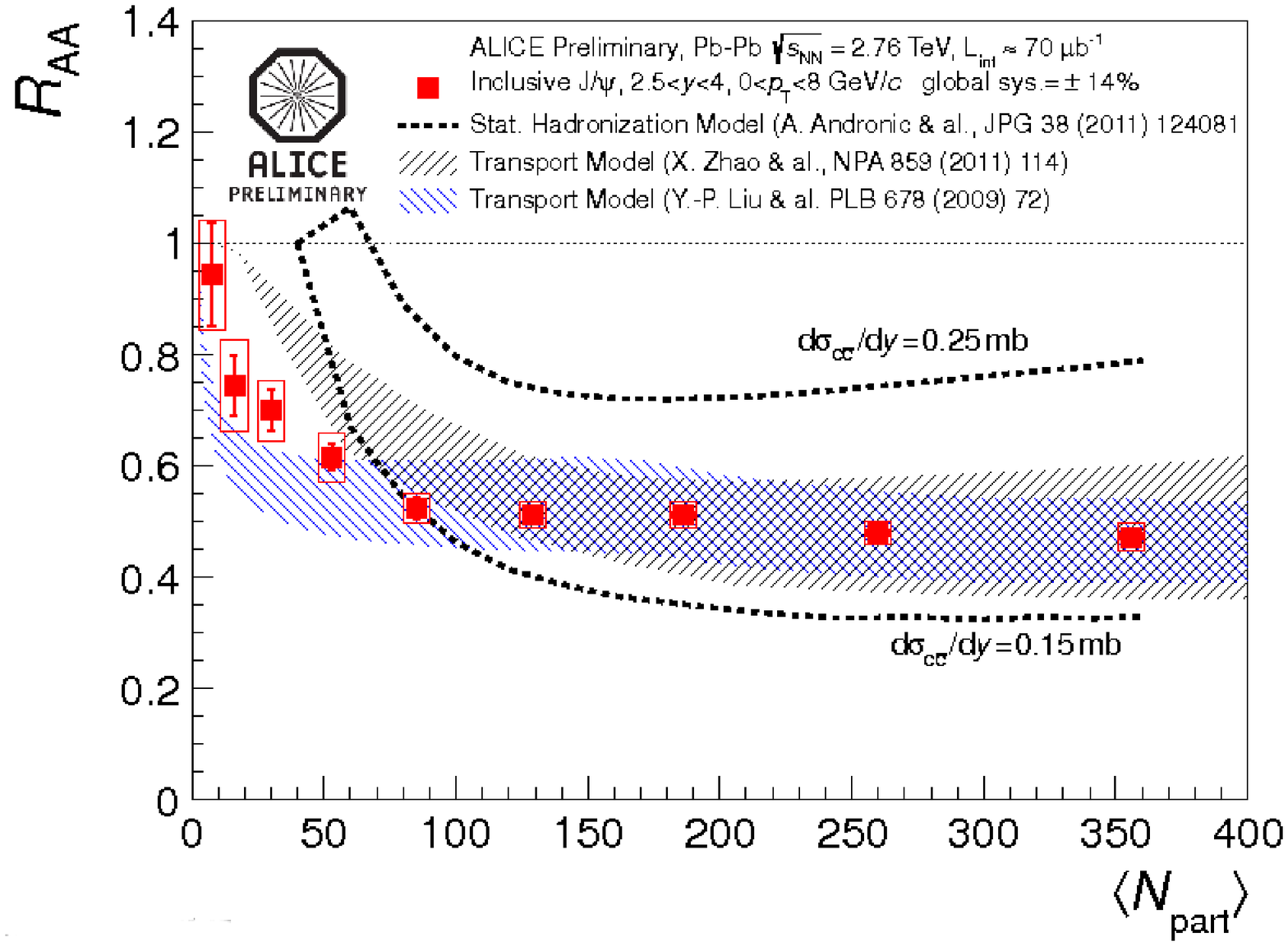}
\caption{\label{fig7}(Color Online) Inclusive $J/$$\psi$  $R_{\rm AA}$ as a function of the number of 
participating nucleons measured in 
Pb-Pb collisions at $\sqrt{s_{\rm NN}}$ = 2.76 TeV. These results are compared 
with statistical hadronization model~\cite{Andronic:2011yq} along with transport models by 
Zhao {\it et al.}~\cite{Zhao:2011cv} and Liu {\it et al.}~\cite{Liu:2009nb} respectively.}
\end{figure}

The forward rapidity(2.5$<y<$4) quarkonia analysis with dimuons in 
Pb-Pb collisions at $\sqrt{s_{\rm NN}}$ = 2.76 TeV is based on a sample 
corresponding to an integrated luminosity of 70 $\mu$b$^{-1}$ which corresponds to 17.7M triggered events.
The invariant mass spectrum of $\mu^{+}\mu^{-}$  candidates for 0$\%$-90$\%$ most central events is shown in Fig.~\ref{fig1}. 
The signal is extracted by performing a combined fit for the background and signal contribution. 
The background is described by a Gaussian with a width($\sigma$) which varies
as a function of the mass value. For the signal description a modified Crystal Ball 
function was used which is a convolution of a Gaussian and power law functions 
that can fit the tails of the measured signal.

The nuclear modification factor($R_{\rm AA}$) 
allow us to quantify the medium effects on the production of $J/$$\psi$. 
$R_{\rm AA}$ gives the deviation in  $J/$$\psi$ yield
in AA collisions relative to the scaled 
(according to the number of binary nucleon-nucleon collisions) 
yield of $J/$$\psi$ in pp collisions. 
The raw yield has been corrected for acceptance and efficiency.

Figure~\ref{fig2} shows the inclusive $J/$$\psi$  $R_{\rm AA}$ for mid- and forward rapidity 
for Pb-Pb collisions at $\sqrt{s_{\rm NN}}$ = 2.76 TeV, as a function of  
the average number of nucleons participating to the collision ($N_{\rm part}$) which has been 
calculated using the Glauber model~\cite{Aamodt:2010cz}.
Its important to note that in ALICE for both the di-muon and di-electron channel the  $J/$$\psi$ 
production can be measured down to $p_{T}$ = 0  GeV/$c$.
The di-muon data sample analysed and shown here are for nine centrality classes from  0$\%$-10$\%$(central collisions) 
to 80$\%$-90$\%$(peripheral collisions). For the dielectron analysis the three centrality classes of 
0$\%$-10$\%$, 10$\%$-40$\%$ and 40$\%$-80$\%$ are shown here.
At forward rapidity the centrality integrated value of $R_{\rm AA}^{0\%-90\%}=0.497\pm0.006({\rm stat.})\pm0.078({\rm syst.})$ 
indicate $J/$$\psi$ suppression, improving the statistical significance of the  
$R_{\rm AA}$ measurements obtained with 2010 data~\cite{alicePbPb2010}.
The centrality integrated mid-rapidity value is 
$R_{\rm AA}^{0\%-80\%}=0.66\pm0.10({\rm stat.})\pm0.24({\rm syst.})$. 
The systematic error is dominated by the pp reference both for  mid- and forward rapidity 
measurements.

Comparison with the RHIC measurements at $\sqrt{s_{\rm NN}}=200$~GeV~\cite{Adare:2011yf} 
from the PHENIX experiment shown in Fig.~\ref{fig3} exhibit that the 
inclusive $J/$$\psi$ $R_{\rm AA}$ at 2.76~TeV in the ALICE forward rapidity 
region are higher than that measured at 200~GeV in the rapidity domain of $1.2 < |y| < 2.2$. 
High-$p_{T}$ $J/$$\psi$ $R_{\rm AA}$  measured in ALICE for 5$<$$p_{T}$$<$8 GeV/$c$ range is compared with 
CMS measurements at central rapidity($|y|<$0.9) for  6.5$<$$p_{T}$$<$30 GeV/$c$ in Fig.~\ref{fig4}.
For ALICE the $J/$$\psi$ $R_{\rm AA}$ at most central collisions is $\sim$ 0.35 and the 
integrated value of $R_{\rm AA}^{0\%-90\%}=0.384\pm0.014({\rm stat.})\pm0.074({\rm syst.})$. 
We observe from Fig.~\ref{fig4} that selecting high-$p_{T}$ drives 
down the $R_{\rm AA}$.

Figure~\ref{fig5} shows $R_{\rm AA}$ as a function of $p_{T}$ as measured by 
ALICE(red) results are compared to the CMS data~\cite{Chatrchyan:2012np}(blue) as
well as results from the PHENIX~\cite{Adare:2011yf} (black) at a lower 
collision energy. The $R_{\rm AA}$ is decreasing from 0.6 for low 
$p_{T}$ to about 0.4 at higher $p_{T}$. The 
CMS results (0$\%$-100$\%$ centrality, $1.6 < |y| < 2.4$, $p_{T}$ $>$ 6.5 GeV/$c$) are in agreement with
the ALICE measurements (0$\%$-90$\%$ centrality, 2.5$<$y$<$4, $p_{T}$ $>$ 0) in the overlapping transverse momentum range 
whereas the lower energy results from PHENIX (0-20$\%$ centrality, $1.2 < |y| < 2.2$) show a significantly smaller  $R_{AA}$. 
Figure~\ref{fig6} shows the  $p_{T}$ dependence of $R_{AA}$ obtained in the most central (0$\%$-20$\%$) 
and most peripheral (40$\%$-90$\%$) centrality classes. The $J/$$\psi$ suppression pattern depends on 
the centrality of the collisions. In the most central collisions (0$\%$-20$\%$), the suppression 
increases with the transverse momentum of the $J/$$\psi$. For the peripheral 
collisions (40$\%$-90$\%$), the $p_{T}$ dependence of the $R_{\rm AA}$ is weaker 
and compatible with a flat behaviour. Thus, for $p_{T}$ $>$ 3 GeV/$c$ the $J/$$\psi$ 
suppression is larger in central collisions.

At forward rapidity, the non-prompt $J/$$\psi$  was
measured by the LHCb collaboration to be about 10$\%$ in pp collisions at
$\sqrt{s}$= 7 TeV~\cite{Aaij:2011jh} in the $p_{T}$ range of the present analysis. 
Neglecting the shadowing effects and considering the scaling of beauty production with 
the number of binary nucleon-nucleon collisions, the prompt $J/$$\psi$ $R_{AA}$ is estimated to be(upper limit), 
11$\%$ smaller than the inclusive measurement. While estimating the influence of non-prompt $J/$$\psi$ 
as a function of $p_{T}$ and $y$ on the inclusive $R_{\rm AA}$ results, 
the LHCb measurement at $\sqrt{s}$ = 2.76 TeV is interpolated using CDF and CMS data. 
$J/$$\psi$  from beauty hadrons have a negligible 
influence on the present measurements while assuming a range of energy
loss for the b-quarks from $R_{\rm AA}$(b) = 0.2 to $R_{\rm AA}$(b) = 1.

Figure~\ref{fig7} compares the ALICE results with results from a statistical hadronization 
model~\cite{Andronic:2011yq}, as well as two different transport models~\cite{Zhao:2011cv,Liu:2009nb}. 
The models describe the data within uncertainties for $N_{\rm part}$ larger than 70.
The Statistical Hadronization Model~\cite{Andronic:2011yq} assumes deconfinement and a thermal equilibration of the
bulk of the $c\bar{c}$ pairs. Then the charmonium production occurs at the phase boundary by statistical hadronization
of charm quarks. The two transport model from Zhao {\it et al.}~\cite{Zhao:2011cv} and Liu {\it et al.}~\cite{Liu:2009nb} 
differ mostly in the rate equation controlling the $J/$$\psi$  dissociation and regeneration. 
Both the transport model calculations are shown as a band which connects the results obtained with (lower limit) 
and without (higher limit) shadowing, which can be interpreted as the uncertainty of the prediction.
The model from Zhao {\it et al.} incorporates a simple shadowing estimate leading to a 30$\%$ suppression for the 
most central Pb-Pb collisions assuming the charm cross-section d$\sigma_{c\bar{c}}$/d$y$ $\approx$  0.5 mb at $y$ = 3.25. 
The $J/$$\psi$  from beauty hadrons is estimated at 10$\%$ and no quenching is assumed.
The model from Liu {\it et al.} has shadowing from EKS98 and a smaller charm cross-section 
d$\sigma_{c\bar{c}}$/d$y$ $\approx$  0.38 mb is used.

\begin{figure}
\includegraphics[scale=0.39]{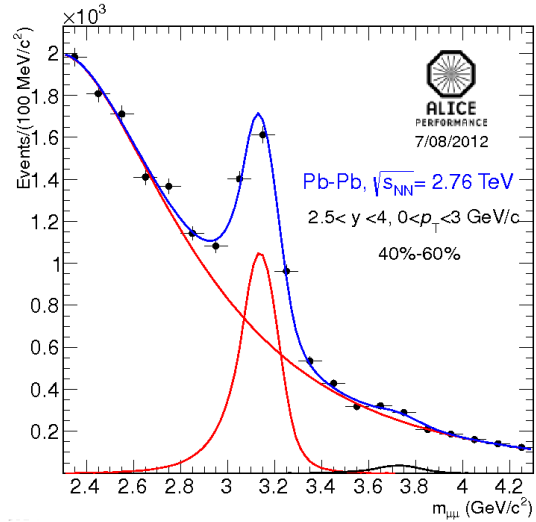}
\caption{\label{fig8}(Color Online) Opposite sign dimuon invariant mass distribution for the 40$\%$-60$\%$ most central collisions and 0$<$$p_{T}$$<$3 GeV/$c$ range in 2.5$<y<$4.}
\end{figure}

\begin{figure}
\includegraphics[scale=0.365]{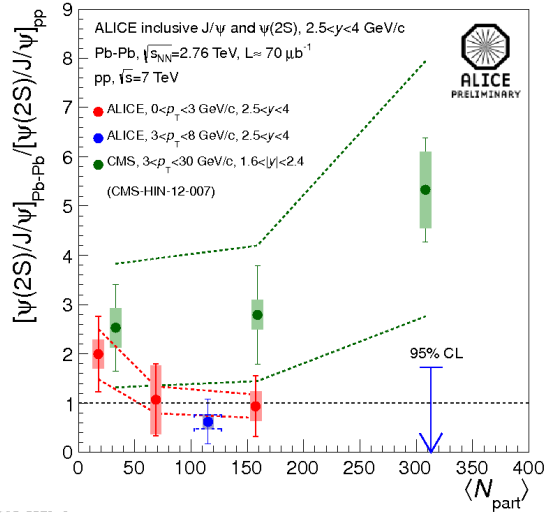}
\caption{\label{fig9}(Color Online) ALICE results for $(N_{\psi(2S)}/N_{J/\psi})_{PbPb}/(N_{\psi(2S)}/N_{J/\psi})_{pp}$ obtained 
in 0$<$$p_{T}$$<$3 and 3$<$$p_{T}$$<$8 GeV/$c$ for Pb-Pb collisions at $\sqrt{s_{\rm NN}}$ = 2.76 TeV recorded in 2011. 
ALICE results are compared with CMS~\cite{cms2S} values for 3$<$$p_{T}$$<$30 GeV/$c$ and 1.6$<y<$2.4.}
\end{figure}

\subsection{$\psi(2S)$}

The study of $\psi(2S)$ provides an interesting comparison with $J/$$\psi$ 
production~\cite{Gupta:1992cd,Chaudhuri:2003ju}.
In particular, a substantial fraction of the $J/$$\psi$'s
is known to originate from $\psi(2S)$  and $\chi_{c}$ decays~\cite{Antoniazzi:1992iv}. 
Examining the ratio of $\psi(2S)$ to $J/$$\psi$ production can render a further insight on the mechanism
affecting quarkonium in a hot and dense medium which was studied in detail 
at SPS energies~\cite{Baglin:1994ui,Alessandro:2006ju}. In a scenario, in which quarkonium states
are suppressed by a Debye screening mechanism, the $\psi(2S)$ meson melts at lower temperatures,
being a less bound state with respect to the $J/$$\psi$~\cite{Gupta:1992cd}. 
A reduction of the $\psi(2S)$ nuclear modification factor,
with respect to the $J/$$\psi$ can infer on this sequential melting~\cite{Gupta:1992cd}. However this picture
might be complicated by charmonium production via recombination mechanisms at LHC, which can affect both
the $J/$$\psi$ and the $\psi(2S)$. Production via recombination mechanisms should 
contribute mostly in the low-$p_{T}$ region which can be studied by the ALICE experiment.

The invariant mass spectrum of $\mu^{+}\mu^{-}$ candidates for 40$\%$-60$\%$ most central events 
and 0$<$$p_{T}$$<$3 GeV/$c$ is shown in Fig.~\ref{fig8}.
Figure~\ref{fig9} shows the double ratio $(N_{\psi(2S)}/N_{J/\psi})_{PbPb}/(N_{\psi(2S)}/N_{J/\psi})_{pp}$, which 
compares the ratio of $\psi(2S)$ over $J/$$\psi$ yields in Pb-Pb and pp. The pp results used 
here are from $\sqrt{s}$ = 7 TeV data-sample due to the low integrated luminosity at $\sqrt{s}$ = 2.76 TeV. 
ALICE results which do not show large enhancement in central collisions are obtained 
in 0$<$$p_{T}$$<$3 GeV/$c$ and 3$<$$p_{T}$$<$8 GeV/$c$ for Pb-Pb collisions at $\sqrt{s_{\rm NN}}$ = 2.76 TeV
and compared with CMS~\cite{cms2S} values for 3$<$$p_{T}$$<$30 GeV/$c$ and 1.6$<$y$<$2.4 as 
a function of centrality. SPS measurements have indicated that $\psi(2S)$ 
is more suppressed than $J/$$\psi$~\cite{Alessandro:2006ju}.

\section{Summary and Outlook}

Nuclear modification factor $R_{\rm AA}$ of $J/$$\psi$ have been measured in 
Pb-Pb collisions at $\sqrt{s_{\rm NN}}$ = 2.76 TeV at forward rapidity and compared with 
mid-rapidity. It was shown that $R_{\rm AA}$ decreases with
increasing $p_{T}$ and towards larger rapidity. Several models that take into account a recombination
of $c\bar{c}$-pairs are compared with the ALICE results.

The inclusive $J/$$\psi$ production in Pb-Pb collisions at $\sqrt{s_{\rm NN}}=2.76$~TeV
show less suppression compared to PHENIX results at RHIC energies.
The regeneration mechanism explain these experimental results. But a 
more quantitative calculation requires the study of cold nuclear matter.
There the roles of the suppression and regeneration mechanisms will be 
disentangled by quantifying the initial shadowing effect with data 
from the forthcoming p-Pb run at LHC, planned in early 2013.

Studying the bottomonia ($\Upsilon(1S)$, $\Upsilon(2S)$, $\Upsilon(3S)$) states 
and their suppression are important at LHC energies. Since the $\Upsilon$ is smaller 
and has larger binding energy than the $J/$$\psi$, the states will provide 
valuable information complementary to the study of charmonium. 
The smaller $b\bar{b}$ rate results in a lower probability for production by coalescence.
In terms of AA collisions, the $\Upsilon$ is expected to dissociate at a
higher temperature~\cite{Strickland:2011aa,Song:2011nu,Strickland:2011mw} than all the other quarkonium states,
thus proving to be a more effective thermometer of the system~\cite{:2012fr,Chatrchyan:2011pe}.







\end{document}